\documentclass[final,1p,times]{elsarticle}
\usepackage{amssymb}
\usepackage{subfigure}
\usepackage{graphicx}% Include figure files
\usepackage{dcolumn,color}% Align table columns on decimal point
\usepackage{bm}
\usepackage{epstopdf}
\usepackage{subeqnarray}
\usepackage{epsfig}
\usepackage{color,framed}
\usepackage{amsmath}
\usepackage[table]{xcolor}
\usepackage{cases}
\usepackage{subfigure}
\usepackage{mathrsfs}
\begin{document}

\begin{frontmatter}



\title{Kinks in higher derivative scalar field theory}

\author[label1]{Yuan Zhong}
\author[label1]{Rong-Zhen Guo}
\author[label1]{Chun-E Fu}
\author[label2]{Yu-Xiao Liu\corref{cor1}}
\cortext[cor1]{Corresponding author: liuyx@lzu.edu.cn}

\address[label1]{School of Science, Xi'an Jiaotong University, Xi'an 710049, China}
\address[label2]{Institute of Theoretical Physics \& Research Center of Gravitation, Lanzhou University, Lanzhou 730000, China}

\begin{abstract}
We study static kink configurations in a type of two-dimensional higher derivative scalar field theory whose Lagrangian contains second-order derivative terms of the field. The linear fluctuation around arbitrary static kink solutions is analyzed. We find that, the linear spectrum can be described by a supersymmetric quantum mechanics problem, and the criteria for stable static solutions can be given analytically. We also construct a superpotential formalism for finding analytical static kink solutions. Using this formalism we first reproduce some existed solutions and then offer a new solution. The properties of our solution is studied and compared with those preexisted. We also show the possibility in constructing twinlike model in the higher derivative theory, and give the consistency conditions for twinlike models corresponding to the canonical scalar field theory.
\end{abstract}


\end{frontmatter}

% \linenumbers

\section{Introduction}
Kink is the simplest topological defect. It exists in nonlinear scalar field theories with at least two degenerated vacua, and has been studied in many branches of physics~\cite{Vachaspati2006}. In the early study of kinks, the scalar field theory is assumed to be canonical, and its Lagrangian can be written as $\mathcal{L}_0=X-V(\phi)$. Here $X\equiv-\frac12(\partial_\mu\phi)^2$ represents the standard kinetic term. In this simple theory, kink solutions can be obtained by choosing suitable scalar potentials. Two well-known solutions are the $Z_2$ symmetric $\phi^4$ kink and the periodic sine-Gordon kink~\cite{Vachaspati2006}.

These two solutions have many differences. For example, the sine-Gordon model supports the interesting breather solution, while in the $\phi^4$ model one can only find an approximation to the breather solution called oscillon, which emits radiation~\cite{CampbellSchonfeldWingate1983,SegurKruskal1987}. The oscillon solutions can also be found numerically in many other models with higher-order polynomial scalar potentials, such as $\phi^6$~\cite{GaniKudryavtsevLizunova2014} and $\phi^8$ potentials~\cite{GaniLenskyLizunova2015}. Another difference between the two models lies in their linear perturbation spectra. The $\phi^4$ model has two bound states: a zero mode and a massive excitation, but the sine-Gordon model only has a zero mode. The massive excitation of the $\phi^4$ model leads to the bounce windows when two kinks collide~\cite{CampbellSchonfeldWingate1983}.

Recently, with the development of cosmology, many noncanonical scalar field theories were proposed~\cite{CapozzielloLaurentis2012,CliftonFerreiraPadillaSkordis2012,JoyceJainKhouryTrodden2015,NojiriOdintsovOikonomou2017,BelenchiaLetiziaLiberatiCasola2018}. In a typical noncanonical scalar field theory (dubbed as the K-field theory), the Lagrangian is assumed to be an arbitrary function of $\phi$ and $X$. This theory was originally applied in cosmology~\cite{Armendariz-PiconDamourMukhanov1999,GarrigaMukhanov1999,Armendariz-PiconMukhanovSteinhardt2001}, and later was used to construct kink solutions either in two-dimensional Minkowski space~\cite{AdamSanchez-GuillenWereszczynski2007,BazeiaLosanoMenezesOliveira2007,BazeiaLosanoMenezes2008,AlmeidaBazeiaLosanoMenezes2013,ZhongLiu2014}, or in five-dimensional warped space~\cite{BazeiaGomesLosanoMenezes2009,BazeiaLobLosanoMenezes2013,ZhongLiu2013,ZhongLiuZhao2014a,ZhongLiu2015}.
In order to find analytical kink solutions in K-field theory, one can use the superpotential method, which rewrites the original second-order differential equations into some first-order ones by introducing the so-called superpotential (see for example Refs.~\cite{ZhongLiu2014,BazeiaGomesLosanoMenezes2009,BazeiaLobLosanoMenezes2013,ZhongLiuZhao2014a}). The linear perturbation of static K-field kinks was systematically investigated in Refs.~\cite{ZhongLiu2014,ZhongLiu2013}.

The Lagrangian of K-field contains only $\phi$ and its first-order derivative $X$. It is natural to ask can we extend the K-field Lagrangian by adding the second-order derivatives of $\phi$, such as $Y\equiv\partial_\mu\partial^\mu\phi$? In fact, this is not a new idea. The study of higher-order derivative theories dates back to the nineteenth century~\cite{Ostrogradsky1850}, and the result is now concluded as the Ostrogradski's theorem, which states that all the Hamiltonians of non-degenerate higher time
derivative theory suffer from linear instabilities (for more details see Refs.~\cite{Woodard2007,Woodard2015}). This instability can be avoided in some special models whose equations of motion are second order despite the presence of higher-order derivatives in the Lagrangians. A well-known example is the Galileon field~\cite{NicolisRattazziTrincherini2009}, whose Lagrangian takes the following form in $1+1$ dimensions:
\begin{equation}
\mathcal{L}=\partial_\mu\phi\partial^\mu\phi+\alpha \partial_\mu\phi\partial^\mu\phi \square \phi.
\end{equation}
Soliton solutions in Galileon field theory have been explored in Refs.~\cite{EndlichHinterbichlerHuiNicolisEtAl2011,MasoumiXiao2012,Zhou2012,PadillaSaffinZhou2011a,BazeiaLosanoSantos2014}. Especially, by using a zero-mode argument, the authors of Ref.~\cite{EndlichHinterbichlerHuiNicolisEtAl2011} showed that the Galileon field cannot give rise to static solitonic solutions.

Thus, in order to find static kink solutions in higher derivative theory, one needs to extend the Galileon theory. In four-dimensional curved space-time, the most general scalar-tensor theory with second-order equations of motion is the Horndeski theory~\cite{Horndeski1974}.
But later it was realized that second-order equation is not mandatory for avoiding the Ostrogradski's instability. The Ostrogradski's instability can also be eliminated by introducing constraints~\cite{ChenFasielloLimTolley2013,ChenLim2014}, or in multifield models~\cite{RhamMatas2016}. Nowadays, the most general extensions to the Horndeski's theory are the so-called degenerate higher-order scalar-tensor (DHOST) theories~\cite{GleyzesLangloisPiazzaVernizzi2015,GleyzesLangloisPiazzaVernizzi2015a,LangloisSaitoYamauchiNoui2018}.

It is interesting to study the static kink solutions in various kinds of higher derivative scalar field theories, and see how the higher derivative terms affect the well-known properties of the canonical kinks. Some successful examples can be found in~\cite{BazeiaLobaoMenezes2015a,CarrilloMasoumiSolomonTrodden2016}. Both works considered the so-called generalized Galileon theory~\cite{DeffayetDeserEsposito-Farese2009}, and the corresponding Lagrangian in two-dimensional Minkowski space reads
\begin{equation}
\label{eqBazeia}
\mathcal{L}=f_1(\phi, X)+f_2(\phi, X) Y.
\end{equation}
In this paper, we extend the works of Refs.~\cite{BazeiaLobaoMenezes2015a,CarrilloMasoumiSolomonTrodden2016} to a model with the following Lagrangian
\begin{equation}
\label{actionLyphi}
\mathcal{L}=\mathcal{L}(\phi, X, Y).
\end{equation}
This lagrangian can be regarded as a simple subclass of the DHOST theories, and the corresponding equation of motion reads
\begin{equation}
\label{EqEOMGen}
{\mathcal{L}_\phi } + {\partial ^\mu }({\mathcal{L}_X}{\partial _\mu }\phi ) + {\partial ^\mu }{\partial _\mu }{\mathcal{L}_Y}=0.
\end{equation}
Here we have defined ${\mathcal{L}_\phi } \equiv\frac{\partial \mathcal{L}}{\partial \phi},$ and so on.
Our aim is to find static kink solutions in a two-dimensional Minkowski space-time with line element
$ds^2=-dt^2+dx^2$.

The paper is organized as follows. In the next section, we firstly give a general discussion on the linear stability of an arbitrary static solution of Eq.~\eqref{EqEOMGen}. We will show that under some conditions, the perturbation equation can be written as a factorizable Schr\"odinger-like equation, which ensures the stability of the solution. In Sec. \ref{SecSup}, we construct the superpotential formalism corresponding to our model. This formalism is powerful in finding kink solutions. As examples, we will apply it to reproduce some of the solutions of \cite{BazeiaLobaoMenezes2015a}, and then give our own solution. After that, we will consider, in Sec.~\ref{SecTwin}, a constrained system. The constraint forces the equation of the higher derivative theory taking the same form as the one of the canonical theory. In this case, nonlinear terms of $Y$ are allowed if some conditions were satisfied. We will also derive the equations that $\mathcal{L}(\phi,X,Y)$ has to satisfy in order to be a twinlike model of $\mathcal{L}_0$. Our results will be summarized in Sec.~\ref{SecSum}.

\section{Linear stability of static configuration}
Suppose we have obtained a static solution $\phi_c(x)$ of Eq.~\eqref{EqEOMGen}, it is important to consider the behavior of a small perturbation $\delta \phi(t,x)=\sum_{n=0}^{\infty} \psi_n(x)e^{i\omega_n t}$ around $\phi_c(x)$. The spectrum of $\omega_n$ can be obtained by solving the linear perturbation equation. Obviously, a real $\omega_n$ corresponds to a stable oscillation $\delta \phi(t,x)$, with frequency $\omega$, around $\phi_c(x)$ . While, an imaginary $\omega_n$ corresponds to an exponentially growing perturbation, and would destroy the original configuration $\phi_c(x)$. Therefore, when $\omega_n^2\geq 0$ holds for all $n$, we say that the static configuration $\phi_c(x)$ is stable against small perturbation. Otherwise, $\phi_c(x)$ is unstable.

In Ref.~\cite{BazeiaLobaoMenezes2015a}, Bazeia et al analyzed the linear perturbation of a model described by the Lagrangian \eqref{eqBazeia}. In this section, we will consider the linearization of static solution of model~\eqref{actionLyphi}. To derive the linear equation of $\delta \phi(t,x)$, one can expand the action around $\phi_c(x)$ up to the second order of the perturbation:
\begin{eqnarray}\label{EqdeltaTwoL}
  {\delta ^{(2)}}\mathcal{L} &=& {\mathcal{L}_X}{\delta ^{(2)}}X
  + \frac{1}{2}{\mathcal{L}_{\phi \phi }}{(\delta \phi )^2}
  + \frac{1}{2}{\mathcal{L}_{XX}}{({\delta ^{(1)}}X)^2} \nonumber \\
  &+& \frac{1}{2}{\mathcal{L}_{YY}}{(\delta Y)^2}
   + {\mathcal{L}_{\phi X}}\delta \phi {\delta ^{(1)}}X
   + {\mathcal{L}_{XY}}{\delta ^{(1)}}X\delta Y\nonumber \\
   &+& {\mathcal{L}_{\phi Y}}\delta \phi \delta Y
   +\mathcal{O}(\delta \phi^3).
\end{eqnarray}
Here we have defined the following quantities:
\begin{eqnarray}
{\delta ^{(1)}}X &=&  - (\partial ^\mu \delta \phi) (\partial_\mu\phi)
= - \delta \phi '\phi ',\\
{\delta ^{(2)}}X &=& - \frac{1}{2}({\partial ^\mu }\delta\phi )({\partial _\mu }\delta\phi ),\\
\delta Y &=& {\partial ^\mu }{\partial _\mu }\delta \phi.
\end{eqnarray}
Obviously, the term $\frac{1}{2}{\mathcal{L}_{YY}}{(\delta Y)^2}$ inevitably leads to fourth-order derivatives terms in the linear perturbation equation. For simplicity, in this work we only consider the case with ${\mathcal{L}_{YY}}=0$, so that the linear perturbation equation is second order. But it does not mean that $\mathcal{L}$ can only contain a linear term of $Y$. As we will see in Sec.~\ref{SecTwin}, sometimes, ${\mathcal{L}_{YY}}$ is vanished after the background equation of motion is considered. In such case, nontrivial higher-order terms of $Y$ are allowed, and do not change the final statements of this section.

At a first glance, the term ${\mathcal{L}_{XY}}{\delta ^{(1)}}X\delta Y=-{\mathcal{L}_{XY}}({\partial ^\mu }\delta \phi )({\partial _\mu }\phi ){\partial ^\nu }{\partial _\nu }\delta \phi$ would also lead to a third-order derivative of $\delta\phi$ after an integration by parts. However, the higher-order derivative terms can be eliminated in the following sense:
\begin{eqnarray}
&&{\mathcal{L}_{XY}}{\delta ^{(1)}}X\delta Y=
 - \frac{1}{2}{\mathcal{L}_{XY}}({\partial ^\mu }\delta \phi )({\partial _\mu }\phi )\square\delta \phi - \frac{1}{2}{\mathcal{L}_{XY}}({\partial ^\mu }\delta \phi )({\partial _\mu }\phi )\square\delta \phi\nonumber\\
&=&\frac{1}{2}\delta \phi {\partial ^\mu }({\mathcal{L}_{XY}}{\partial _\mu }\phi \square\delta \phi )
   - \frac{1}{2}\delta \phi \square({\mathcal{L}_{XY}}{\partial _\mu }\phi {\partial ^\mu }\delta \phi )+\partial_\mu(\cdots),
\end{eqnarray}
where the last term in the second line is a total derivative term. Obviously, the terms that contain $\partial^\mu{\partial ^\nu }{\partial _\nu }\delta \phi$ are canceled.

In the end, for a static background kink configuration, the quadratic Lagrangian density of $\delta\phi$ reads
\begin{eqnarray}
 && {\delta ^{(2)}}\mathcal{L} =\frac{1}{2}({\mathcal{L}_X} + \mathcal{L}_{XY}'\phi ' + {\mathcal{L}_{XY}}\phi '' + 2{\mathcal{L}_{\phi Y}}) \delta \phi \square \delta \phi\nonumber \\
  &+& \frac{1}{2}{\mathcal{L}_{\phi \phi }}{(\delta \phi )^2}
   + \delta \phi \delta \phi '(\frac{1}{2}\mathcal{L}_X'
   - \frac{1}{2}\mathcal{L}_{XX}'\phi '^2
   - {\mathcal{L}_{XX}}\phi '\phi ''\nonumber \\
  &-&  {\mathcal{L}_{\phi X}}\phi '
  -\frac{1}{2}\mathcal{L}_{XY}''\phi ' - \mathcal{L}_{XY}'\phi '' - \frac{1}{2}{\mathcal{L}_{XY}}\phi ''') \nonumber \\
   &-& \delta \phi \delta \phi ''(\frac{1}{2}{\mathcal{L}_{XX}}\phi '^2 + \mathcal{L}_{XY}'\phi ' + {\mathcal{L}_{XY}}\phi '').
\end{eqnarray}
By defining the following variables
\begin{eqnarray}
\label{EqVariableG}
\mathcal{G} &=& \delta \phi\sqrt{\xi},\\
z &=& \phi '\sqrt {\xi},\\
\xi &\equiv&{\mathcal{L}_X} + \mathcal{L}_{XY}^\prime \phi ' + {\mathcal{L}_{XY}}\phi '' + 2{\mathcal{L}_{\phi Y}},\\
\gamma &=& 1-\frac{\phi '^2 }{z^2}\left(2 \mathcal{L}_{XY}' \phi '+\mathcal{L}_{XX} \phi '^2+2 \mathcal{L}_{XY} \phi ''\right),
\end{eqnarray}
the quadratic Lagrangian density can be simplified as
\begin{equation}
 {\delta ^{(2)}}\mathcal{L} = \frac{1}2\left\{
   -\mathcal{G}\partial_t^2\mathcal{G} +\mathcal{V}(x)\mathcal{G}^2
+\gamma\mathcal{G}\mathcal{G}''\right\},
\end{equation}
where
\begin{equation}
\mathcal{V}(x)=-\gamma \frac{z''}{z}-\frac{z'}{z}\gamma '-\frac{1}{2}\gamma ''.
\end{equation}

If $\gamma>0$, we can introduce a new spatial coordinate $x^{\ast}$, such that
\begin{equation}
\label{RWcoord}
\frac{dx^{\ast}}{dx}\equiv\gamma^{-1/2},
\end{equation}
and define
\begin{equation}
\hat{\mathcal{G}}=\frac{1}{\sqrt{2}}\gamma^{1/4}\mathcal{G}.
\end{equation}
Then, the quadratic action can be written as
\begin{equation}
\label{G2}
 {\delta ^{(2)}}{S_{\hat{\mathcal{G}}}} = \int dtdx^{\ast}\hat{\mathcal{G}}
 \left\{-\partial_t^2\hat{\mathcal{G}}+ \ddot{\hat{\mathcal{G}}} -\frac{\ddot{\theta}}{\theta}\hat{\mathcal{G}}
\right\},
\end{equation}
where
\begin{equation}
\theta \equiv\gamma^{1/4}z,
\end{equation}
and an over dot represents the derivative with respect to $x^\ast$.

From the quadratic action of $\hat{\mathcal{G}}$, we know that for
\begin{equation}
\label{criterion}
\xi>0,\quad \gamma>0,
\end{equation}
the normal mode of the linear perturbation satisfies a Schr\"odinger-like equation
\begin{equation}
\label{schroScalar}
-\ddot{\hat{\mathcal{G}}} +\frac{\ddot{\theta}}{\theta}\hat{\mathcal{G}}
=-\partial_t^2\hat{\mathcal{G}}.
\end{equation}
The normal mode of the scalar perturbation can be expanded as
\begin{equation}
\hat{\mathcal{G}}=\sum_{n=0}^{\infty} \psi_n(x^\ast)e^{i\omega_n t},
\end{equation}
and the equation for $\psi_n(x^\ast)$ reads
\begin{equation}
H \psi_n=-\ddot{\psi}_n+V_{\textrm{eff}} \psi_n=\omega_n^2 \psi_n,
\end{equation}
where $H=-\frac{d^2}{dx^{\ast 2}}+V_{\textrm{eff}}$ is the Hamiltonian of linear perturbation, and $V_{\textrm{eff}}=\ddot{\theta}/{\theta}$ is the effective potential.
It is easy to check that this Hamiltonian can be factorized as
\begin{equation}
\label{hatG}
H=\mathcal{A}\mathcal{A}^\dagger,
\end{equation}
where
\begin{equation}
\label{Adagger}
\mathcal{A}=\frac{d}{dx^{\ast}}+\frac{\dot{\theta}}{\theta},\quad
\mathcal{A}^\dagger=-\frac{d}{dx^{\ast}}+\frac{\dot{\theta}}{\theta}.
\end{equation}
According to the study of supersymmetic quantum mechanics~\cite{CooperKhareSukhatme1995}, a system with a factorizable Hamiltonian has two important properties:
\begin{enumerate}
  \item $\omega_n$ are semipositive definite, namely, $\omega_n\geq 0$. The zero mode ($\omega_0=0$) of $H$ reads
\begin{eqnarray}
\label{eqZeroMode}
\psi_0=c  \theta(x^{\ast}),
\end{eqnarray} where $c$ is the normalization constant.
  \item One can construct a partner Hamiltonian
\begin{eqnarray}
H_{-}&=&\mathcal{A}^\dagger \mathcal{A}
=-\frac{d^2}{dx^{\ast 2}}+V_{\textrm{sup}},
\end{eqnarray}
where $V_{\textrm{sup}}\equiv\theta\left(\theta^{-1}\right)^{..}$. $H_{-}$ and $H$ share the same spectrum, except for the zero mode $\psi_0$.
\end{enumerate}
The first property ensures the stability of any static background solution that satisfies the inequalities \eqref{criterion}. While the superpartner potential $V_{\textrm{sup}}$ offers us an alternative way to analyze the spectrum of the linear perturbation. As we will see in Sec. \ref{SecSup}, sometimes $V_{\textrm{sup}}$ is more convenient for us to discern, at least qualitatively, the properties of the linear spectrum.

\section{The superpotential method}
\label{SecSup}
As we have mentioned in the previous section, our arguments on the linear stability are valid only when $\mathcal{L}_{YY}= 0$. This can be satisfied in two cases. In the first case, the Lagrangian contains at most the linear term of $Y$, which is nothing but the generalized Galileon model given in \eqref{eqBazeia}.
In the other case, $\mathcal{L}_{YY}$ vanished when the equation of motion is substituted.
In this section, we will consider the first case, and give the corresponding superpotential formalism for the first time.

From Eq.~\eqref{EqEOMGen} we know that a static solution $\phi(x)$ satisfies
\begin{equation}
\label{EqEOMGenStatic}
{\mathcal{L}_\phi } + ({\mathcal{L}_X}\phi' )' +\mathcal{L}_Y''=0.
\end{equation}
It is not difficult to show (by integration) that
\begin{equation}
\label{EqL}
-\rho=\mathcal{L}=  2{\mathcal{L}_X}X- \mathcal{L}_Y'\phi ' + {\mathcal{L}_Y}\phi ''+ C.
\end{equation}
Here $\rho$ is the energy density of the kink and $C$ is an integral constant. To ensure a kink has finite energy, we require that $\lim_{|x|\to\infty}\rho=0$, which requires $C=0$.

Since $\phi(x)$ is odd and $\mathcal{L}(x)=-\rho(x)$ is even, we require $\mathcal{L}_Y$ to be an odd function of $x$ such that all terms in Eq.~\eqref{EqL} have the same parity. A simple choice is to set $f_2=g(\phi) F(X)$, where $F(X)$ is an arbitrary function of $X$, and $g(\phi)$ is an odd function of $\phi$. In Ref.~\cite{BazeiaLobaoMenezes2015a} the authors considered a simple case with $F(X)\propto X$ and $g=\phi$. In the present work, we consider the following subclass model:
\begin{eqnarray}
\label{eqf1}
f_1(\phi,X)&=&K(X)-V(\phi),\\
\label{eqf2}
f_2(\phi,X)&=& g(\phi) F(X).
\end{eqnarray}
Here $K(X), V(\phi), U(X)$ are functions of the corresponding arguments.

The superpotential $W(\phi)$ is introduced in the normal way~\cite{ZhongLiu2014}

\begin{equation}
\label{eqFirst}
\phi'=W(\phi).
\end{equation}

Plugging Eqs.~\eqref{eqf1}-\eqref{eqf2} into Eq.~\eqref{EqL}, we find that the scalar potential $V$ and the energy density $\rho$ take the following expressions:
\begin{eqnarray}
\label{EqV}
V&=&K-2X\left( Fg_{\phi}+K_X \right) ,\\
\label{Eqrho2}
\rho &=&-2X\left( Fg_{\phi}+K_X \right) -FgY.
\end{eqnarray}
The stability criteria \eqref{criterion} read
\begin{eqnarray}
\label{eqCriteriaOne}
\xi &=&2gY\left( XF_{\text{XX}}+F_X \right) -2XF_Xg_{\phi}+2Fg_{\phi}+K_X>0,\\
\label{eqCriteriaTwo}
\gamma & =&[2\left( XF_X+F \right) g_{\phi}+2XK_{\text{XX}}+K_X]/\xi>0.
\end{eqnarray}

By taking suitable superpotentials, analytical kink solutions can be easily obtained via solving Eq.~\eqref{eqFirst}. For instance, by taking
\begin{equation}
W(\phi)=k v \cos \left(\frac \phi v\right),
\end{equation}
one obtains the sine-Gordon kink
\begin{equation}
 \phi(x)=
   v \arcsin(\tanh(k x)),
\end{equation}
while by taking
\begin{equation}
\label{superPot}
W(\phi)=k v \left(1-\left(\frac \phi v\right)^2\right),
\end{equation}
one obtains
\begin{equation}
 {\phi(x)=}
   v\tanh(kx).
\end{equation}
Here $k$ and $v$ are two constants with dimensions. But for simplicity, we will always take $k=v=1$ in this paper.

Note that in Eqs.~\eqref{EqV} and \eqref{Eqrho2}, $V(\phi)$ and $\rho(x)$ have been expressed as functions of $\phi, X$ and $Y$. They can be given explicitly by simply using the relations $X=-\phi'^2/2=-W^2/2$ and $Y=\phi''=W_\phi W$.
\subsection{Deriving some existed solutions}
To apply the above superpotential formalism, let us first derive the solutions given in Ref.~\cite{BazeiaLobaoMenezes2015a}. For the first model, let us consider the following Lagrangian:
\begin{equation}
\label{Eqmodel1}
\mathcal{L}=X-V(\phi)-b \phi X Y.
\end{equation}
where $b>0$ is a positive constant.
For this model, Eqs.~\eqref{EqV}-\eqref{eqCriteriaTwo} read
\begin{eqnarray}
V&=&X\left( 2bX-1 \right),\\
\rho &=&X\left( 2bX+bY\phi -2 \right),\\
\xi &=&1-2bY\phi>0,\\
\gamma \xi &=&1-4bX>0.
\end{eqnarray}
By taking the superpotential given in Eq.~\eqref{superPot}, one immediately obtains the first kink solution of Ref.~\cite{BazeiaLobaoMenezes2015a}, whose scalar potential and energy density read
\begin{eqnarray}
V&=&\frac{1}{2}\left( 1-\phi ^2 \right) ^2\left( 1+b\left( 1-\phi ^2 \right) ^2 \right),\\
\rho &=&S^4 \left( 1+b S^2 -\frac{1}{2}b S^4 \right),
\end{eqnarray}
where $S=\textrm{sech}(x)$.

Finally, from the definitions of $X$ and $Y$, one can show that for any kink we always have $\phi Y\leq0$ and $X<0$, and therefore, this solution are linearly stable. Also note that in this case, the second order derivative of the energy density is $\rho''(x=0)=-2(2+b)$. That means $\rho$ always peaks at $x=0$ for any $b>0$. No energy density splitting in this case.

Another kink solution of Ref.~\cite{BazeiaLobaoMenezes2015a} has the following Lagrangian:
\begin{equation}
\mathcal{L}=X+bX^2-V(\phi)-\frac 32 b \phi X Y.
\end{equation}
Since the coefficients of $X^2$ and $\phi X Y$ terms are tuned, the corresponding expressions are concise:
\begin{eqnarray}
V&=&-X,\\
\rho &=&-\frac{1}{2}X\left( 2bX-3bY\phi +4 \right) ,\\
\label{eqXi}
\xi &=&1+2 b X-3 b Y \phi,\\
\gamma  &=&1/\xi.
\end{eqnarray}
Interestingly, by using Eq.~\eqref{superPot} one obtains a $\phi^4$ potential
\begin{eqnarray}
V&=&\frac{1}{2}\left( \phi ^2-1 \right)^2,
\end{eqnarray}
and the following energy density:
\begin{eqnarray}
\rho=S^4 \left(1+\frac{3 b}{2}S^2-\frac{7 b}{4}S^4\right).
\end{eqnarray}
The independence of the scalar potential on the deviation parameter $b$, makes this model unambiguous in discussing the effects of the noncanonical kinetic terms.

However, from Eq.~\eqref{eqXi} we know that the positivity of $\xi$ depends on the value of $b$. To see this, we use the kink solution and rewrite Eq.~\eqref{eqXi} as
\begin{eqnarray}
\xi(x)=1+6bS^2-7bS^4.
\end{eqnarray}
Then by solving $\xi'(x)=0$, we know that there are only three extreme points:
\begin{eqnarray}
x_0=0,\quad x_{\pm}=\frac{1}{2} \ln \left(\frac{1}{3} \left(11\pm 4 \sqrt{7}\right)\right).
\end{eqnarray}
To tell which point is the maximum, we need to calculate $\xi''$ at these points, and the results are
\begin{eqnarray}
\xi''(x_0)=16 b,\quad \xi''(x_{\pm})=-\frac{288}{49}b.
\end{eqnarray}
Obviously, $x_0$ and $x_{\pm}$ are the minimum (maximum) and the maxima (minima), respectively, if $b>0$ ($b<0$).

For $b>0$, the range of $\xi(x)$ is
\begin{eqnarray}
1- b=\xi(x_0) \leq\xi(x)\leq \xi(x_{\pm})=1+\frac{9}{7}b ,
\end{eqnarray}
and the stability criterion $\xi>0$ equivalents to $\xi(x_{0})>0$ or $b<1$. In this case, the parameter can only take value in a rather narrow range $0<b<1$. Nevertheless, a nontrivial effect, namely, the splitting of energy density can be found at $x=0$ when $b>4/5$. In order to split, the energy density has to change from a maximum to a local minimum, so $x=0$ must be an inflection point, such that $\rho''(0)=5b-4=0$. The solution is simply $b=4/5$.

Now, let us turn to the case with $b<0$, which is omitted by the authors of Ref.~\cite{BazeiaLobaoMenezes2015a}. In this case, the range of $\xi(x)$ becomes
\begin{eqnarray}
1+\frac{9}{7}b=\xi(x_{\pm}) \leq\xi(x)\leq \xi(x_0)= 1- b,
\end{eqnarray}
and stability of the solution requires $-\frac79<b<0$. Within this parameter space, there is no energy density splitting, because $\rho''(0)=5b-4$ is always negative.

In addition to the above two kink solutions, Ref.~\cite{BazeiaLobaoMenezes2015a} also given a compact solution as well as a double kink solution, whose energy density splitting is more evident. We will not redo the job here.

\subsection{A new solution}
\begin{figure}
\begin{center}
\includegraphics[width=0.4\textwidth]{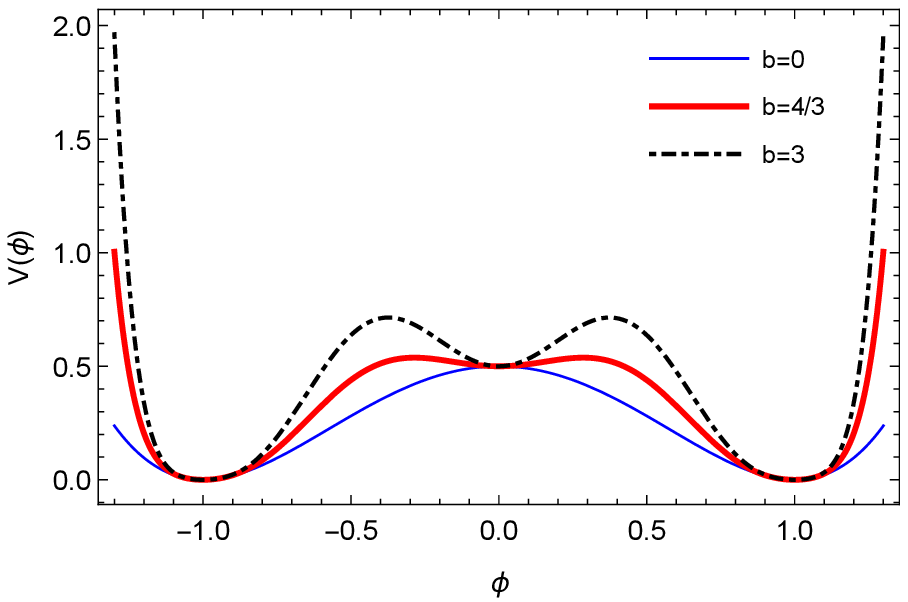}
\includegraphics[width=0.4\textwidth]{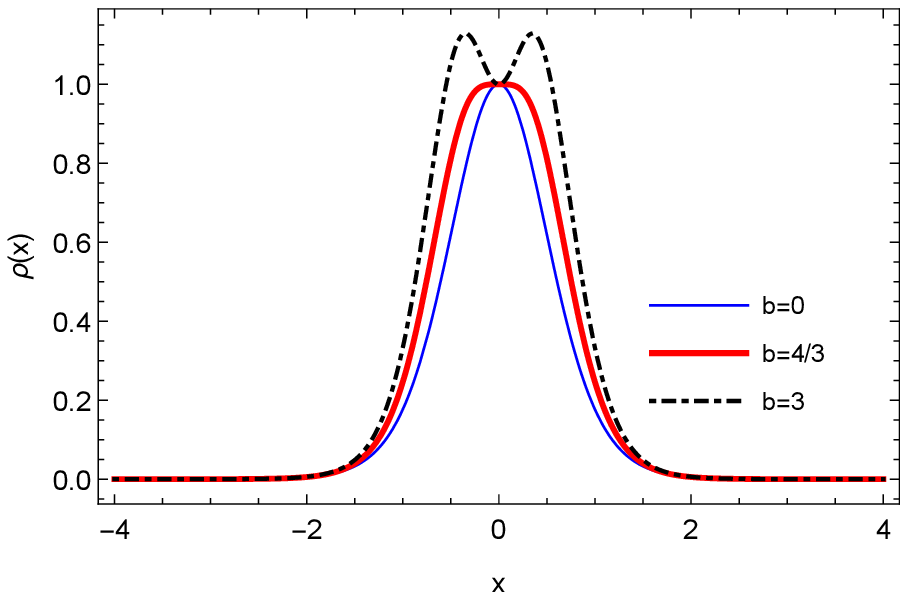}
\caption{The scalar potential and the energy density correspond to Eqs.~\eqref{OurV} and \eqref{OurRho}, respectively. The energy density splits when $b>4/3$.}
\label{figOurRho}
\end{center}
\end{figure}

With the help of superpotential method, one can construct, in principle, infinite models that support stable kink solutions. Instead of writing down a complicated model, we would like to modify the model in Eq.~\eqref{Eqmodel1} as the following one:
\begin{equation}
\label{Model3}
\mathcal{L}=X-V(\phi)-b \phi^3 X Y,\quad b>0.
\end{equation}
Using the superpotential given in Eq.~\eqref{superPot} we get
\begin{eqnarray}
\label{OurV}
V&=&\frac{1}{2} \left(1-\phi ^2\right)^2 \left(3 b \phi ^2 \left(1-\phi ^2\right)^2+1\right),\\
\label{OurRho}
\rho &=&S^4 \left(1+b S^2-\frac{b S^4}{2}-\frac{b S^6}{2}\right).
\end{eqnarray}
In this case $\rho''(x=0)=3 b-4$, which implicates that when $b>4/3$ the energy density begins to split (see Fig.~\ref{figOurRho}).

Now, let us move to the discussion of the linear perturbation. For our model,
\begin{eqnarray}
\xi &=&1-2bY\phi^3,\\
\gamma \xi &=&1-12bX\phi^2.
\end{eqnarray}
As we have mentioned previously, a kink configuration always satisfies $\phi Y\leq0$ and $X<0$. So, for $b>0$ our solution always meets the requirement from the stability criteria $\xi>0,~\gamma>0$.

\begin{figure}
\begin{center}
\includegraphics[width=0.34\textwidth]{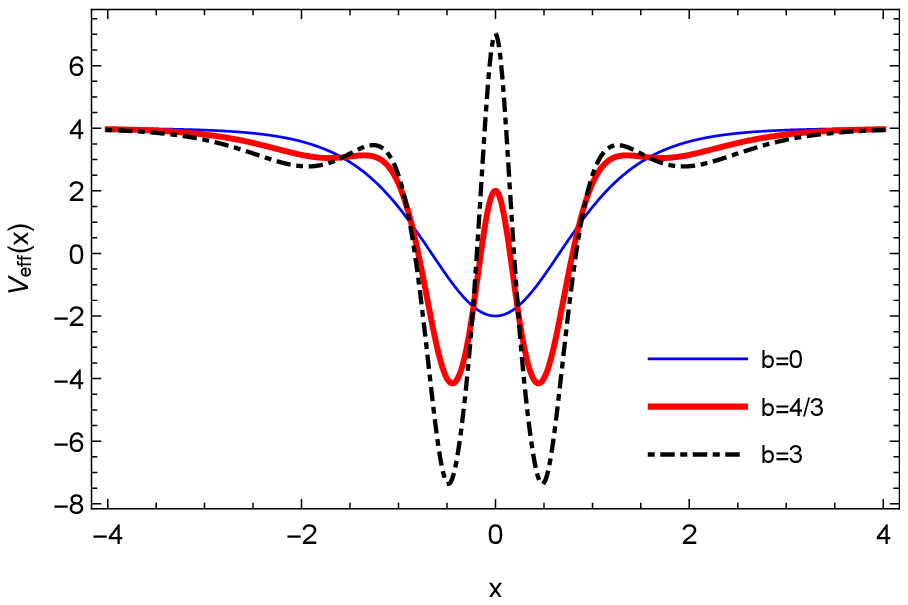}
\includegraphics[width=0.34\textwidth]{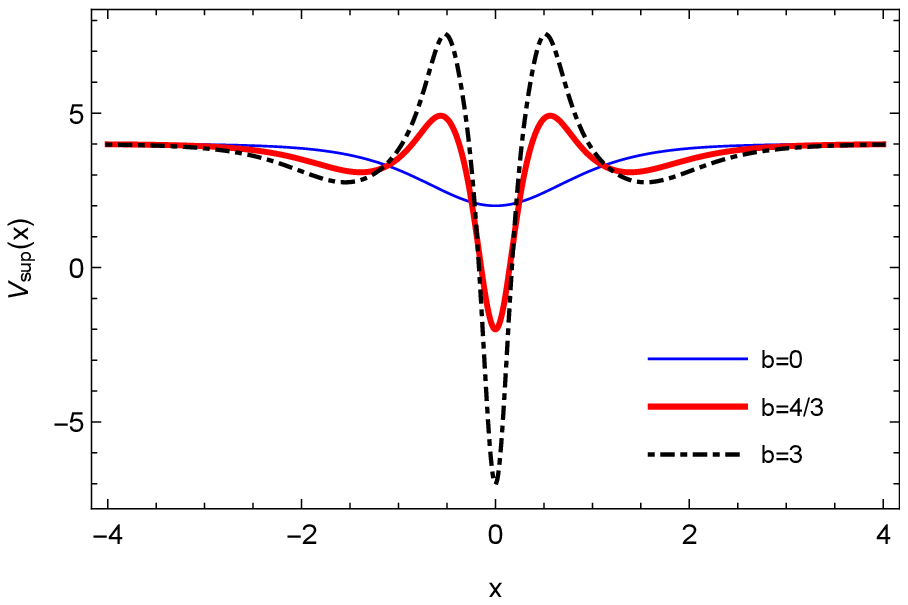}
\caption{The effective potentials $V_{\textrm{eff}}$ and $V_{\textrm{sup}}$ correspond to Eqs.~\eqref{OurV} and \eqref{OurRho}, respectively. The energy density splits when $b>4/3$.}
\label{FigVeff}
\end{center}
\end{figure}

Instead of giving a quantitative calculation of the linear spectrum, which is impossible without using numerical method, we would like to discuss qualitatively what would be different if the deviation parameter $b$ is large enough. As can be seen from Fig.~\ref{FigVeff}, with $b$ increases, an inner structure appears in the effective potential of the linear perturbation. It has been reported previously that inner structure might leads to quasilocalized perturbation modes~\cite{GuoLiuZhaoChen2012,XieYangZhao2013,CruzSousaMalufAlmeida2014}. Indeed, for large $b$, the superpartner potential $V_{\textrm{sup}}(x)$ becomes a volcano potential, which might support metastable states.

\begin{figure}
\begin{center}
\includegraphics[width=0.35\textwidth]{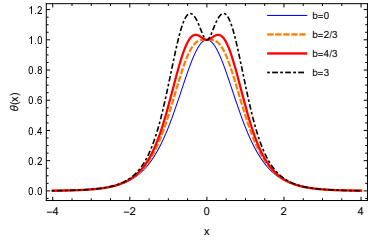}
\caption{Plots of the zero mode $\theta(x)$, which begins to split when $b>2/3$.}
\label{FigTheta}
\end{center}
\end{figure}

Now, let us now turn to the zero mode of the pertubration. As we have shown in Eq.~\eqref{eqZeroMode}, the zero model is proportional to $\theta\equiv\gamma^{1/4}z$. In general, it is impossible to give an analytical expression of $\theta$ in the $x^\ast$ coordinate, because the coordinate transformation \eqref{RWcoord} is so complicated. But, in the $x$ coordinate it is straightforward to use the definition and write out $\theta(x)$. For our solution
\begin{eqnarray}
\theta(x)=S^2 \sqrt[4]{\left(6 b \left(1-S^2\right) S^4+1\right) \left(4 b S^2 \left(1-S^2\right)^2+1\right)}.
\end{eqnarray}
It is easy to show that at $x=0$, $\theta$ and its derivatives are
\begin{eqnarray}
\theta(0)=1,\quad \theta'(0)=0, \quad \theta''(0)=3b-2.
\end{eqnarray}
Thus, as $b>2/3$, $x=0$ varies from a global maximum to a local minimum of $\theta(x)$, and the zero mode splits (see Fig.~\ref{FigTheta}).

In this section, we have shown the power of the first-order formalism in constructing analytically solvable kink models. In addition to reproduce two of the kink solutions given in Ref.~\cite{BazeiaLobaoMenezes2015a}, we also offer a new solution. It is worth to note that the model described by Eq.~\eqref{Eqmodel1} and the one by Eq.~\eqref{Model3} can be written uniformly as follows
\begin{equation}
\mathcal{L}=X-V(\phi)-b \phi^{(2n+1)} X Y,\quad b>0.
\end{equation}
For integer $n$, there is a qualitative difference between models with $n=0$ and $n>0$, namely, the splitting of energy density. As we have mentioned previously, for $n=0$ the energy density does not split. While for $n=1$, Fig.~\ref{figOurRho} shows that $\rho$ begins to split when the deviation parameter $b$ is large enough. One can also show that the same thing happens for models with $n>1$.

In addition to the superpotential method, there exists another interesting method to generate kink solutions, namely, the twinlike model method.

\section{Twinlike model method}
\label{SecTwin}
Twinlike models are defined as two apparently different models that share the same field configuration and the same energy density~\cite{ZhongLiu2015,AndrewsLewandowskiTroddenWesley2010,AdamQueiruga2011,BazeiaDantasGomesLosanoMenezes2011,BazeiaMenezes2011,AdamQueiruga2012,BazeiaDantas2012,BazeiaHoraMenezes2012,BazeiaLobMenezes2012,GomesMenezesNobregaSimas2013,BazeiaLobaoLosanoMenezes2014,GomesMenezesNobregaSimas2014,ZhongFuLiu2016}. Take the canonical model
\begin{equation}
\label{EqMGModel}
\mathcal{L}_0=X-V(\phi)
\end{equation} as an example. It has the following Dirac-Born-Infeld (DBI) type twinlike model
\begin{eqnarray}
\label{DBIMin}
\mathcal{L}_{\textrm{DBI}}=U(\phi)\sqrt{1-X}+\epsilon_0,
\end{eqnarray}
if
\begin{eqnarray}
U(\phi)=-2\sqrt{1+V(\phi)}, \quad \epsilon_0=2.
\end{eqnarray}
The DBI model is only a special case of the K-field model for which $\mathcal{L}=\mathcal{L}(\phi,X)$, and in fact one may construct infinite K-field type twinlike models for $\mathcal{L}_0$ (see~\cite{AndrewsLewandowskiTroddenWesley2010,AdamQueiruga2011,BazeiaDantasGomesLosanoMenezes2011}). The aim of this section is to construct some twinlike models for $\mathcal{L}_0$  in the higher derivative theory $\mathcal{L}=\mathcal{L}(\phi,X,Y)$.

The starting point is the equation of motion \eqref{EqL}:
\begin{equation}
\label{EOM}
\mathcal{L}- 2{\mathcal{L}_X}X+ \mathcal{L}_Y'\phi ' - {\mathcal{L}_Y}\phi ''=0.
\end{equation}
For the canonical field $\mathcal{L}=\mathcal{L}_0$, the equation of motion and the energy density are
\begin{equation}
X=-V,
\end{equation}
and
\begin{equation}
\rho_0=-\mathcal{L}_0=-2X=2V,
\end{equation}
respectively.

Now let us move to the general model $\mathcal{L}=\mathcal{L}(\phi,X,Y)$. To be a twinlike model of the canonical one, $\mathcal{L}$ must satisfy the following condition:
\begin{equation}
\label{EqTwinCond}
\mathcal{L}|=2X=-2V.
\end{equation}
Here the symbol ``$|$" represents taking the on-shell condition $X=-V$.
The one-shell condition ensures that the noncanonical and the canonical models satisfy the same equation of motion, and Eq.~\eqref{EqTwinCond} comes from the requirement of the same energy density, because $\rho=-\mathcal{L}$.
But this requirement must be consistent with the equation of motion \eqref{EOM}. By plugging Eq.~\eqref{EqTwinCond} into Eq.~\eqref{EOM} we obtain
\begin{equation}
2X - 2{{\cal L}_X}X - 2{{\cal L}_{Y\phi }}X + 2{{\cal L}_{YX}}XY - {{\cal L}_Y}Y +{{\cal L}_{YY}}\phi '\phi ''' = 0.
\end{equation}
Since $\mathcal{L}(\phi,X,Y)$ is independent of $\phi'''$, we may conclude that the consistency conditions are
\begin{eqnarray}
\label{eqTwin1}
 (2X - 2{{\cal L}_X}X - 2{{\cal L}_{Y\phi }}X + 2{{\cal L}_{YX}}XY - {{\cal L}_Y}Y)| &=&0,\\
 \label{eqTwin2}
{\cal L}_{YY}|&=&0.
\end{eqnarray}
Obviously, for the K-field $\mathcal{L}=\mathcal{L}(\phi,X)$ the above conditions is simply ${{\cal L}_X}|=1$. One can easily proof that the Lagrangian of the DBI model in Eq.~\eqref{DBIMin} satisfies
$\mathcal{L}_{\textrm{DBI}}|=2X=-2V$ and $\partial\mathcal{L}_{\textrm{DBI}}/\partial X|=1$,
and therefore, is a twinlike model of $\mathcal{L}_0$.

It is not difficult to construct the twinlike model corresponds to $\mathcal{L}_0$ in the higher derivative theory. One of the infinite Lagrangians that satisfy eq.~\eqref{EqTwinCond} and the consistency conditions \eqref{eqTwin1}-\eqref{eqTwin2} is
\begin{equation}
\label{eqNonCan}
\mathcal{L}=X-V+b\phi (X+V)^2  Y ^{2n+1},\quad b>0,
\end{equation}
where $n\geq 0$ is a nonnegative integer.
By solving this model, one would get $\phi(x)=\phi_0(x)$ and $\rho(x)=\rho_0(x)$\footnote{The solution $\phi_0(x)$ is completely determined by the scalar potential $V(\phi)$. We have known many solvable potentials: $\phi^4$, sine-Gordon, and so on.}.  But it does not mean that $\mathcal{L}$ and $\mathcal{L}_0$ are the same model written in two different ways, the essential difference between these two models is that for canonical model $\partial_{X,X}\mathcal{L}_{0}|=0$, while for its higher-order derivative twin
\begin{equation}
 \partial_{X,X}\mathcal{L}|=2b \phi Y^{2n+1}\neq 0.
\end{equation}
As a result, $\xi| =1$, $z| = \phi '$, $\gamma| =1+4 b X Y \phi $,
\begin{equation}
\theta| =(\gamma^{1/4}z)|={(1-4b X Y)}^{1/4}\phi '.
\end{equation}
That means the twinlike models have different zero mode configurations.

It is worth to mention that this model allows nonlinear terms of $Y$ provided that Eq.~\eqref{eqTwin2} is satisfied. Therefore, we can regard model~\eqref{eqNonCan} as a constrained higher-order derivative system, which has been considered as one alternative to go beyond the Horndeski theory~\cite{ChenFasielloLimTolley2013,ChenLim2014}.

\section{Summary}
\label{SecSum}
In this paper, we considered a type of higher-order derivative scalar field theory whose Lagrangian takes the form $\mathcal{L}=\mathcal{L}(\phi,X,Y)$. We derived the quadratic action of the linear perturbation around arbitrary static background solution. For simplicity we assume $\mathcal{L}_{YY}$ is vanished either because $\mathcal{L}$ only contains the linear term of $Y$, or as a consequence of the background equation. If further the Lagrangian satisfies two inequalities \eqref{criterion}, then the normal mode of the scalar perturbation satisfies a Schr\"odinger-like equation, and the corresponding Hamiltonian can be factorized into a form often seen in supersymmetric quantum mechanics. For a factorizable Hamiltonian the spectrum is semi-positive, thus the corresponding solution is linear stable.

We also constructed a superpotential formalism for finding analytical static kink solutions in the generalized Galileon theory. As applications of our formalism, we first reproduced two of the solutions reported in Ref.~\cite{BazeiaLobaoMenezes2015a}, and then gave our own solution. For our solution, both the energy density and the scalar zero mode split when the deviation parameter $b$ is larger enough. Besides, the shape of the superpartner potential $V_{\textrm{sup}}(x)$ implies that metastable resonant states might appear for large $b$. It is not difficult to construct more kink solutions with the superpotential formalism.

Then we explored another possibility where $\mathcal{L}_{YY}$ vanishes when the background equation is considered. In this case, nonlinear terms of $Y$ are allowed, and will not ruin our arguments on the linear stability. As an example, we consider a constrained higher-order derivative system, whose equation of motion is assumed to be the same as the one of the canonical system $\mathcal{L}_0=X-V$. If in addition to the equation of motion, we also assume the two different systems possess identical energy densities, then these two systems are called as twinlike models. In order to be a twin model of $\mathcal{L}_0$, the Lagrangian of the higher-order derivative $\mathcal{L}$ must satisfies the consistency equations~\eqref{eqTwin1}-\eqref{eqTwin2}. By giving an example Lagrangian \eqref{eqNonCan}, we state that twinlike models are essentially different and they can be distinguished by their linear spectra.

\section*{Acknowledgments}
This work was supported by the National Natural Science Foundation of
China (Grant Numbers 11605127, 11522541, 11405121, and 11375075), and by China Postdoctoral Science Foundation (Grant No. 2016M592770).

%\bibliographystyle{model1a-num-names}
%\bibliography{E:/360YunPan/jabref/library/articles/bib/articles}% Produces the bibliography via BibTeX.
%\bibliography{/Users/zhongyuan/jabref/library/articles/bib/articles}% Produces the bibliography via BibTeX.






% \end{thebibliography}

\end{document}